\documentclass[aps,amssymb,showpacs]{revtex4}
\usepackage{graphicx}
\usepackage{color}
\usepackage{ulem}

\begin{document}

\title{ $J/\psi$ Transverse Momentum Distribution in High Energy Nuclear Collisions}
\author{Yunpeng Liu$^1$}
\author{Zhen Qu$^1$}
\author{Nu Xu$^2$}
\author{Pengfei Zhuang$^1$}
\affiliation{$^1$Physics Department, Tsinghua University, Beijing
100084, China\\ $^2$Nuclear Science Division, Lawrence Berkeley
National Laboratory, Berkeley, California 94720, USA}

\date{\today}

\begin{abstract}
The transverse momentum ($p_t$) dependence of $J/\psi$ production
in heavy ion collisions is investigated in a transport model with
both initial production and continuous regeneration of charmonia.
The competition between the two production mechanisms results in a
$p_t$ suppression in central collisions, the gluon
multi-scattering in the initial stage leads to a high $p_t$
enhancement, and the regeneration populates $J/\psi$s at low $p_t$
region and induces a minimum in $R_{AA}(p_t)$. These three
phenomena are indeed observed in both 200 GeV Cu+Cu and Au+Au
collisions at RHIC energy.
\end{abstract}

\pacs{25.75.-q, 12.38.Mh, 24.85.+p}

\maketitle

\begin{figure}[!hbt]
\centering
\includegraphics[width=0.5\textwidth]{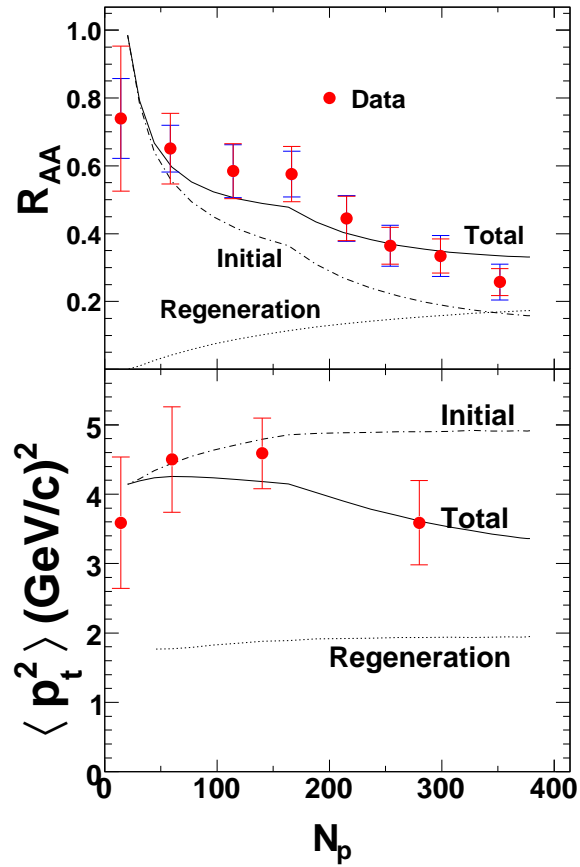}
\vspace{-0.25cm} \caption{The nuclear modification factor $R_{AA}$
and averaged transverse momentum square $\langle p_t^2 \rangle$ as
functions of participant nucleon number $N_p$ in central rapidity
region for Au+Au collisions at RHIC energy. Dot-dashed, dotted and
solid lines represent respectively the calculations with only
initial production, only regeneration and both contributions. The
data are from PHENIX collaboration~\cite{data1,data2}.}
\label{fig1}
\end{figure}

\begin{figure}[!hbt]
\centering
\includegraphics[width=0.5\textwidth]{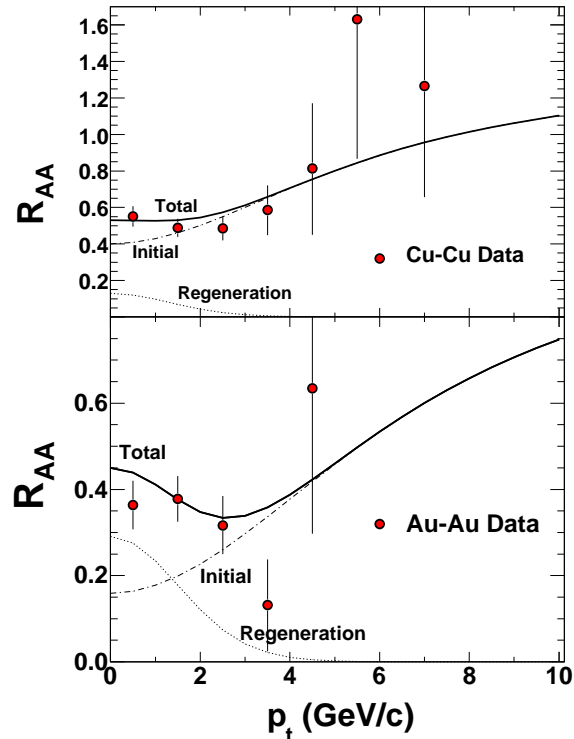}
\vspace{-0.5cm} \caption{The nuclear modification factor $R_{AA}$
as a function of transverse momentum $p_t$ in central rapidity
region for Cu+Cu and Au+Au collisions at RHIC energy. Dot-dashed,
dotted and solid lines represent respectively the calculations
with only initial production, only regeneration and both
contributions. The data are from STAR~\cite{data3} for Cu+Cu and
PHENIX~\cite{data2} for Au+Au.} \label{fig2}
\end{figure}

 The $J/\psi$ suppression in relativistic heavy ion collisions is
widely accepted as a signature of the formation of Quark-Gluon
Plasma (QGP) at finite temperature and density~\cite{matsui}.
Recently, the PHENIX and STAR collaboration presented the
transverse momentum distribution of $J/\psi$ production in
Au+Au~\cite{data1,data2} and Cu+Cu~\cite{data3} collisions at RHIC
energy $\sqrt{s_{NN}}$ = 200 GeV. In particular the new data
display three interesting features (see Figs.\ref{fig1} and
\ref{fig2}): 
\begin{enumerate}
 \item The averaged transverse momentum square $\langle
p_t^2\rangle$ is strongly suppressed in central collisions, which
is very different from the SPS data~\cite{data4} where $\langle
p_t^2\rangle$ gets saturated for central Pb+Pb collisions at
energy $\sqrt{s_{NN}}\approx 17.3$ GeV;
 \item The nuclear modification factor $R_{AA}$ is about 1/2 at low $p_t$ but becomes
around unity at high $p_t$;
 \item There exists a minimum in $R_{AA}$ in low $p_t$ region.
\end{enumerate}
Can these phenomena tell us something new
about the nature of $J/\psi$ production at RHIC energy?

The $p_t$ broadening at SPS energy is generally attributed to the
gluon multi-scattering in the initial state~\cite{hufner1} and the
leakage effect in the final state~\cite{hufner2}. In a pA
collision, a gluon of the proton scatters from target nucleons
before it fuses with a gluon from the target to form a $J/\psi$.
The gluon rescattering in the initial state is treated as a random
walk in transverse momentum and the observed $\langle
p_t^2\rangle$ is predicted to increase linearly with the mean
length of the path of the incident gluon. In an AB collision, both
gluons which fuse to a $J/\psi$ are affected by the rescattering.
The leakage effect on $J/\psi$ production has already been
considered 20 years ago~\cite{matsui,blaizot,karsch1}. The
anomalous suppression inside the QGP is not an instantaneous
process, but takes a certain time. During this time the $J/\psi$s
with high transverse momenta may leak out of the source of the
anomalous suppression. As a consequence, low $p_t$ $J/\psi$s are
absorbed preferentially but high $p_t$ $J/\psi$s can survive.

While charm quark production at SPS energy is expected to be
small, there are more than 10 $c\bar{c}$ pairs produced in a
central Au+Au collision at RHIC energy and probably more than 200
pairs at LHC energy~\cite{gavai}. The uncorrelated charm quarks in
the QGP can be recombined to form $J/\psi$s. Obviously, the
regeneration will enhance the $J/\psi$ yield and alter its
momentum spectrum. The regeneration approach for $J/\psi$
production at RHIC has been widely discussed with different
models, such as thermal production on the hadronizaton
hypersurface according to statistic law~\cite{munzinger}, the
coalescence mechanism~\cite{greco}, and the kinetic
model~\cite{thews} which considers continuous regeneration in a
QGP. Recently, the $J/\psi$ transverse momentum distribution at
RHIC energy was discussed~\cite{zhao} in the frame of a
two-component model~\cite{rapp} which includes both initial
$J/\psi$ production through nucleon-nucleon (NN) interaction and
regeneration.

The medium created in high-energy nuclear collisions evolves
dynamically. In order to extract information about the medium by
analyzing the $J/\psi$ distribution, both the hot and dense medium
and the $J/\psi$ production processes must be treated dynamically.
Since the massive $J/\psi$s are unlikely fully thermalized with
the medium, their phase space distribution is governed by a
transport equation. In the transport approach~\cite{zhuang} which
includes dissociation and regeneration processes in a QGP, the
$p_t$ broadening due to initial gluon rescattering and the leakage
effect in the final state are respectively taken into account
through the initial condition and the free streaming term of the
transport equation. To comprehensively treat the $J/\psi$
distribution in the transport approach, the $J/\psi$ transport
equation is solved together with hydrodynamic equations which
characterize the space-time evolution of the QGP. In this Letter,
we investigate in the transport model the $J/\psi$ transverse
momentum distributions in heavy ion collisions at RHIC and LHC
energies, including the $\langle p_t^2\rangle$ as a function of
centrality and the $R_{AA}$ as a function of $p_t$.

In pp collisions, the $J/\psi$s from the feed-down of $\psi'$ and
$\chi_c$ are respectively about 10$\%$ and 30$\%$ of the total final
$J/\psi$s~\cite{feeddown}. Since $\Psi (=J/\psi,\psi',\chi_c)$ is
heavy, the distribution function $f_\Psi({\bf p}_t,{\bf
x}_t,\tau|{\bf b})$ in central rapidity region and transverse phase
space $({\bf p}_t,{\bf x}_t)$ at time $\tau$ and fixed impact
parameter ${\bf b}$ is controlled by a classical Boltzmann-type
transport equation~\cite{zhuang}
\begin{equation}
\label{transport}
\partial f_\Psi/\partial \tau +{\bf
v}_\Psi\cdot{\bf \nabla}f_\Psi = -\alpha_\Psi f_\Psi +\beta_\Psi.
\end{equation}
The second term on the left-hand side arises from free-streaming of
$\Psi$ with transverse velocity ${\bf v}_\Psi = {\bf p}_t/\sqrt{{\bf
p}_t^2+m_\Psi ^2}$, which leads to the leakage effect and is
important for those high momentum charmonia. The suppression and
regeneration in hot medium are reflected in the loss term
$\alpha_\Psi({\bf p}_t,{\bf x}_t,\tau|{\bf b})$ and gain term
$\beta_\Psi({\bf p}_t,{\bf x}_t,\tau|{\bf b})$. Considering only the
gluon dissociation process $g+\Psi\to c+\bar c$, $\alpha_\Psi$ can
be expressed as
\begin{equation}
\label{alpha}
\alpha_\Psi({\bf p}_t,{\bf x}_t,\tau|{\bf b}) =
{1\over 2E_\Psi}\int{d^3{\bf p}_g\over (2\pi)^3
2E_g}W_{g\Psi}^{c\bar c}(s)f_g\left({\bf
p}_g,T,u)\right)\Theta\left(T-T_c\right)/\
\Theta\left(T_d^\Psi-T\right),
\end{equation}
where $E_g$ and $E_\Psi$ are the gluon and charmonium energies,
$f_g=1/(e^{p_g^\mu u_\mu/T}-1)$ is the gluon thermal distribution,
$T({\bf x}_t,\tau|{\bf b})$ and $u({\bf x}_t,\tau|{\bf b})$ are
the local temperature and velocity of the hot medium, and
$W_{g\Psi}^{c\bar c}$ is the transition probability of the gluon
dissociation as a function of $s=(p+p_g)^2$ calculated with
perturbative Coulomb potential~\cite{peskin}. For $J/\psi$, the
dissociation cross section reads
\begin{equation}
 \sigma_{gJ/\psi}^{c\bar c}(\omega)=A_0\frac{(\omega/\epsilon_{J/\psi}-1)^{3/2}}{(\omega/\epsilon_{J/\psi})^5}
\label{sigma}
\end{equation}
with $A_0=(2^{11}\pi/27)(m_c^3\epsilon_{J/\psi})^{-1/2}$, where
$\omega$\ is the gluon energy in the rest frame of $J/\psi$, $m_c$
is the charm quark mass, and $\epsilon_{J/\psi}$ is the binding
energy of $J/\psi$. From what discussed in \cite{weise}, to take
the relativistic effect into consideration and to avoid
non-physics divergence in the regeneration cross section, we
replace the $J/\psi$ binding energy by the gluon threshold energy.

The step function in the numerator of the loss term $\alpha$,
controlled by the critical temperature $T_c$ of deconfinement phase
transition, means that the dissociation happens only in the QGP
phase, namely $\alpha=0$ for $T<T_c$. Since the hadronic phase
occurs later in the evolution of heavy ion collisions when the
density of the system is lower compared to the early hot and dense
period, we have neglected the hadronic dissociation. The step
function in the denominator of $\alpha$, characterized by the
dissociation temperature $T_d^\Psi$, indicates that any $\Psi$ cannot 
survive when the temperature of the QGP is higher than
$T_d^\Psi$, namely $\alpha=\infty$ for $T>T_d^\Psi$. From the
lattice QCD simulation~\cite{asakawa}, the $J/\psi$ spectral
function remains almost the same when the temperature is between
$T_c$\ and $T_d$, but the sharp peak of $J/\psi$ vanishes suddenly
when the temperature reaches $T_d$. The introduction of the
dissociation temperature in $\alpha$ is also consistent with the
idea of sequential charmonium suppression~\cite{matsui,satz}.

The gain term $\beta$ in the transport equation (\ref{transport}),
which is a function of $c$ and $\bar c$ recombination transition
probability $W_{c\bar c}^{g\Psi}$, can be obtained from the lose
term $\alpha$ using detailed balance~\cite{thews,zhuang}. Instead
of the gluon distribution $f_g$ in $\alpha$, the charm quark
distribution $f_c$ in $\beta$ is assumed to be in the form of
\begin{equation}
f_c\left({\bf p}_c,{\bf x}_t,\tau\right|{\bf
b})=\frac{d\sigma_{NN}^{c\bar{c}}}{dy}\Big|_{y=0}\frac{T_A({\bf
x}_t)T_B({\bf x}_t-{\bf b})}{\tau} f_{cp}\left({\bf p}_c|{\bf
b}\right),
\end{equation}
where $d\sigma^{c\bar c}_{NN}/dy|_{y=0}$ is the production cross
section of charm quark pairs at central rapidity region in NN
collisions, and $T_A$ and $T_B$ are the thickness functions of the
two colliding nuclei defined as $T({\bf x}_t)=\lim_{z_1\to
-\infty,z_2\to\infty}T({\bf x}_t,z_1,z_2)$ with $T({\bf
x}_t,z_1,z_2)=\int_{z_1}^{z_2} dz \rho({\bf x}_t,z)$ and $\rho({\bf
r})$ being the Woods-Saxon nuclear density profile. From the PHENIX
data, open charms carry an elliptic flow $v_2\sim 0.1$~\cite{v2}
which means that charm quarks are likely thermalized with the
medium, we take the charm quark momentum distribution $f_{cp}$ as
the Fermi-Dirac function $f_{cp}\left({\bf p}_c, T,u|{\bf
b}\right)\sim 1/(e^{p_c^\mu u_\mu/T}+1)$.

With the known suppression and regeneration terms $\alpha$ and
$\beta$, the transport equation can be solved analytically with
the result~\cite{zhuang}
\begin{eqnarray}
f_\Psi\left({\bf p}_t,{\bf x}_t,\tau|{\bf b}\right) &=&
f_\Psi\left({\bf p}_t,{\bf x}_t-{\bf
v}_\Psi(\tau-\tau_0),\tau_0|{\bf
b}\right)e^{-\int^{\tau}_{\tau_0}d\tau'
\alpha\left({\bf p}_t,{\bf x}_t-{\bf v}_\Psi(\tau-\tau'),\tau'|{\bf b}\right)}\nonumber\\
&&+\int^{\tau}_{\tau_0}d\tau' \beta\left({\bf p}_t,{\bf x}_t-{\bf
v}_\Psi(\tau-\tau'),\tau'|{\bf
b}\right)e^{-\int^{\tau}_{\tau'}d\tau''\alpha\left({\bf p}_t,{\bf
x}_t-{\bf v}_\Psi(\tau-\tau''),\tau''|{\bf b}\right)}.
\label{solution}
\end{eqnarray}
The first and second terms on the right-hand side indicate the
contributions from the initial production and continuous
regeneration, respectively. Both suffer anomalous suppression. The
coordinate shift ${\bf x}_t\to{\bf x}_t-{\bf v}_\Psi\Delta\tau$
reflects the leakage effect during the time period $\Delta\tau$.

Since the collision time for NN interactions at RHIC energy is
about 0.1 fm/c and less than the starting time $\tau_0$ of the
medium evolution which is about 0.5 fm/c, the nuclear absorption
for the initially produced charmonia has ceased before the QGP
evolution and is reflected in the transport process as the initial
distribution $f_\Psi\left({\bf p}_t,{\bf x}_t,\tau_0|{\bf
b}\right)$ of the solution (\ref{solution}). Considering a finite
formation time of charmonia which is about 0.5 fm/c and larger
than the collision time, the nuclear absorption can be safely
neglected for the calculations in heavy ion collisions at high
energy~\cite{cassa}. In this case, the initial distribution can be
written as
\begin{equation}
\label{initial} f_\Psi\left({\bf p}_t,{\bf x}_t,\tau_0|{\bf
b}\right)={5\over 4\pi}{d\sigma_{NN}^\Psi\over dy}\Big|_{y=0}\int
dz_Adz_B\rho_A({\bf x}_t,z_A)\rho_B({\bf x}_t-{\bf b},z_B){1\over
\langle p_t^2\rangle_N}\left(1+{p_t^2\over 4\langle
p_t^2\rangle_N}\right)^{-6}
\end{equation}
with a normalized power-law momentum distribution extracted from
the PHENIX data for pp collisions~\cite{ptpp}, where
$d\sigma_{NN}^\Psi/dy|_{y=0}$ is the charmonium production cross
section at central rapidity region in NN collisions, and $\langle
p_t^2\rangle_N$ is the averaged transverse momentum square after
NN collisions~\cite{hufner1}
\begin{equation}
\label{pt2nn} \langle p_t^2\rangle_N({\bf x}_t,z_A,z_B|{\bf
b})=\langle p_t^2\rangle_{NN}+ a_{gN}\rho_0^{-1}\left(T_A({\bf
x}_t,-\infty,z_A)+ T_B({\bf x}_t-{\bf b},z_B,\infty)\right)
\label{eq:cronin_effect}
\end{equation}
with $\rho_0$ being normal nuclear density. For given values ${\bf
b}$ and ${\bf x}_t$ in the transverse plane, suppose a $\Psi$ is
produced at longitudinal coordinates $z_A$ and $z_B$ in nuclei A and
B, respectively. The two gluons which fuse to form the $\Psi$ carry
transverse momentum from two sources: 1) Intrinsic $p_t$, because
they had been confined to nucleons. The intrinsic part is observable
via $NN\to\Psi$ process and leads to $\langle p_t^2\rangle_{NN}$ in
Eq.(\ref{pt2nn}). 2) Collisional contribution to $p_t$, because in a
nuclear matter, the gluons traverse thickness $T_A({\bf
x}_t,-\infty,z_A)$ and $T_B({\bf x}_t-{\bf b},z_B,\infty)$ of
nuclear matter in A and B, respectively, and acquire additional
transverse momentum via $gN$ collisions. This is the origin of the
second term in Eq.(\ref{pt2nn}). The constant $a_{gN}$ is usually
adjusted to the data for pA collisions.

The local temperature $T({\bf x}_t,\tau|{\bf b})$ and fluid
velocity $u_\mu({\bf x}_t,\tau|{\bf b})$, which appear in the
thermal gluon and charm quark distribution functions and control
the suppression and regeneration region via the two step functions
in $\alpha$ and $\beta$, are determined by the $(2+1)$ dimensional
Bjorken's hydrodynamic equations for the medium evolution,
\begin{eqnarray}
\label{hydro}
\partial_{\tau}E+\nabla\cdot{\bf M} &=& -(E+p)/{\tau}\
,\nonumber\\
\partial_{\tau}M_x+\nabla\cdot(M_x{\bf v}) &=& -M_x/{\tau}-\partial_xp\ ,
\nonumber\\
\partial_{\tau}M_y+\nabla\cdot(M_y{\bf v}) &=& -M_y/{\tau}-\partial_yp \
,\nonumber\\
\partial_{\tau}R+\nabla\cdot(R{\bf v}) &=& -R/{\tau}
\end{eqnarray}
with the definitions $E=(\epsilon+p)\gamma^2-p$, ${\bf
M}=(\epsilon+p)\gamma^2{\bf v}$ and $R=\gamma n$, where $\gamma$
is the Lorentz factor, and $\epsilon, p$ and ${\bf v}$ are the
local energy density, pressure and transverse velocity of QGP.

To close the hydrodynamical equations we need to know the equation
of state of the medium. We follow Ref.\cite{sollfrank} where the
deconfined phase at high temperature is an ideal gas of massless
$u$, $d$ quarks, 150 MeV massed $s$ quarks and gluons, and the
hadron phase at low temperature is an ideal gas of all known
hadrons and resonances with mass up to 2 GeV~\cite{pdg}. There is
a first order phase transition between these two phases. In the
mixed phase, the Maxwell construction is used. The mean field
repulsion parameter and the bag parameter are chosen as $K$=450
MeV fm$^3$~\cite{sollfrank} and $B^{1/4}$=236 MeV to obtain the
critical temperature $T_c=165$ MeV at vanishing baryon number
density. The initial condition of the hydrodynamic equation at
RHIC energy is the same as in Ref.\cite{zhuang}.

Solving the local temperature and fluid velocity from the
hydrodynamic equations and then substituting them into the
transport solution (\ref{solution}) for $J/\psi, \psi'$ and
$\chi_c$, we obtain the distribution function $f_{J/\psi}({\bf
p}_t,{\bf x}_t,\tau_f|{\bf b})$ for the final state $J/\psi$s at
the freeze-out time $\tau_f$. By employing the Cooper-Frye
formula~\cite{cooper} with a longitudinal Hubble-like
fluid~\cite{wong}, any physical observable A for $J/\psi$ can be
estimated by integrating $A\cdot f_{J/\psi}$ over the freeze-out
surface.

We now calculate the nuclear modification factor $R_{AA}$ and
averaged transverse momentum square $\langle p_t^2\rangle$ as
functions of centrality decided by the number of participant
nucleons $N_p$ for $J/\psi$s produced in Cu+Cu and Au+Au
collisions at RHIC energy. The following parameters are used in
our calculations: the charm quark mass $m_c=1.87$ GeV~\cite{rapp},
the charmonium mass $m_{J/\psi}$ =3.6 GeV and
$m_{\psi'}=m_{\chi_c}$ =3.7 GeV, the starting time of hydrodynamic
evolution $\tau_0$=0.6 fm, and the charm quark and charmonium
production cross section at $\sqrt{s}=200$ GeV
$d\sigma_{NN}^{c\bar{c}}/dy|_{y=0}=120\ \mu$b~\cite{ccpp} and
$B_{ll}d\sigma_{NN}^{\Psi}/dy|_{y=0}$ =26.4, 4.4, and 13.2 nb for
$\Psi=J/\psi,\ \psi'$ and $\chi_c$~\cite{ptpp}.

The $\Psi$ binding energy or the gluon threshold energy in hot and
dense medium should be smaller than its value in vacuum. It is
estimated to be less than 220 MeV in the QGP
phase~\cite{karsch2,rapp1}. We will take $\epsilon_{J/\psi}$=150
MeV. From our numerical results, a not very large deviation from
this value does not lead to a sizeable change in $R_{AA}$ and
$\langle p_t^2\rangle$. The quasi-free cross section with a
modified binding energy by the medium effect has been discussed in
detail in Ref.\cite{zhao}. Such a quasi-free process has a larger
width than that of the gluon dissociation with binding energy in
vacuum.

The results of $R_{AA}(N_p)$ and $\langle p_t^2\rangle(N_p)$ at
central rapidity for Au+Au collisions are shown in Fig.\ref{fig1}
and compared with the RHIC data~\cite{data1,data2}. The $R_{AA}$
is defined as the ratio of $J/\psi$s produced in a nuclear
collision to that in a corresponding pp collision, normalized by
the number of binary collisions. When the centrality increases,
the contribution from the initial production drops down
monotonously, due to the suppression in the QGP, while the number
of regenerated $J/\psi$s goes up monotonously, because the number
of charm quarks increases with the collision centrality. From the
experimental data, there exists a flat region at 50$<N_p<$170 with
$R_{AA}\simeq 0.6$, and at $N_p > 170$ the $R_{AA}$ continues to
decrease with increasing centrality. This suppression structure
can be reproduced in our transport approach by choosing the
dissociation temperature $T_d^{\psi'}=T_d^{\chi_c}=T_c$ and
$T_d^{J/\psi}=1.92\ T_c$. After the plateau ends, the space-time
region with temperature $T>T_d^{J/\psi}$ increases with
centrality, and more $J/\psi$s are fully eaten up by the
extremely hot QGP. This is the reason of the further decrease of
$R_{AA}$ at $N_p>170$. Since the QGP with $T>T_d$ exists only in
the early stage of the fireball and lasts for a short time, the
initially produced charmonia experience this stage, while most of
the regenerated charmonia that are created later do not. This is
why the $R_{AA}$ for regenerated $J/\psi$s is much less influenced
by the dissociation temperature. The participant number
fluctuations becomes important to $R_{AA}$ at extremely large
$N_p$, as discussed at SPS energy~\cite{fluctuation}, and may
explain the deviation of our result from the data in very central
collisions.

We also calculated the $R_{AA}$ for Cu+Cu collisions and compared
it with the RHIC data~\cite{data5}. When the colliding energy is
fixed, the temperature of the fireball is mainly controlled by the
participant number, and therefore the result for Cu+Cu is almost
the same as that for semi-central Au+Au collisions with $N_p<110$,
where 110 is the maximum participant number for a Cu+Cu collision.
Since Cu is much lighter than Au, the fireball formed in Cu+Cu
collisions cannot reach the temperature for full $J/\psi$
dissociation.

Most of the produced particles are with low momentum~\cite{data2},
they dominate the centrality dependence of the yield, and the
information carried by high $p_t$\ charmonia is therefore screened
in $R_{AA}(N_p)$. In order to understand the behavior of those
high $p_t$ $J/\psi$s which are more sensitive to the production
and suppression mechanisms, we consider the averaged transverse
momentum square $\langle p_t^2\rangle$ in the bottom panel of
Fig.\ref{fig1}. In our calculation, the values of $\langle
p_t^2\rangle_{NN}=4.14$ (GeV/c)$^2$~\cite{ptpp} and $a_{gN}=0.1$
(GeV/c)$^2$/fm~\cite{zhao} are used for NN collisions, see
eq.(\ref{eq:cronin_effect}). At SPS, the gluon rescattering
parameter was taken as $a_{gN}=0.077$ (GeV/c)$^2$/fm~\cite{data4,
na501}. As one can see, the initial contribution to $\langle
p_t^2\rangle$ increases smoothly at $N_p\leq 170$ and shows a
saturation when $N_p>170$. This behavior is very similar to the
case at SPS energy~\cite{hufner2} where there is almost no
regeneration and the initial production can be considered as the
total result. The regenerated $J/\psi$s are from the thermalized
charm quarks, therefore, their transverse momentum is rather small
and the averaged value is almost centrality independent, in
comparison with the initially produced charmonia, see the dotted
line. Since the regeneration becomes more important in central
collisions, the total $\langle p_t^2\rangle$ is strongly
suppressed at large $N_p$ as a consequence of the competition
between the initial production and regeneration. While almost all
the models with and without regeneration mechanism can describe
the $J/\psi$ yield after at least one parameter is adjusted, the
$p_t$ suppression in central collisions seems to be a signature of
charmonium regeneration in Au+Au collisions at RHIC energy. Again,
the calculation for Cu+Cu collisions is very similar to the result
for Au+Au in the region of $N_p<110$.

To look into the details of $J/\psi$ production and suppression
mechanisms, we now turn to the calculation of $R_{AA}(p_t)$ as a
function of $p_t$ which is discussed in different
models~\cite{gunji,zhao}. The result for central Cu+Cu and Au+Au
collisions at midrapidity is shown and compared with the RHIC data
in Fig.\ref{fig2}. The monotonous increase of $R_{AA}$ with only
initial production comes from three aspects. One is the $p_t$
dependence of the gluon dissociation cross section. The gluon
energy relative to $J/\psi$ in the cross section (\ref{sigma}) is
scaled by $J/\psi$ binding energy which is much smaller than 1
GeV. Therefore, gluons with small energy are more likely to
dissociate a $J/\psi$, or in other words, $J/\psi$s with low
momentum are easy to be eaten up by the hot medium. The second
reason is the leakage effect with which the high momentum
charmonia can escape the anomalous suppression region. The last
one is the $p_t$ broadening happened in the initial state. Gluons
get additional transverse momentum via collisions with participant
nucleons before they fuse to $J/\psi$s. At SPS energy the $J/\psi\
R_{AA}$ has been observed to exceed unity at high $p_t$ which can
be well explained by the gluon multi-scattering~\cite{hufner2}. At
RHIC energy, from our numerical calculation, the gluon
rescattering is essential for the increasing of $R_{AA}$\ in high
$p_t$\ region.

In comparison with the initially produced $J/\psi$s which can
carry high momentum from the hard process, the regenerated
$J/\psi$s from thermalized charm quarks are distributed at low
momentum region. In Fig.\ref{fig2}, these effects are clearly
shown by dot-dashed and dotted lines for initial production and
regeneration, respectively. In the low $p_t$ region, the
competition between the initial production which increases with
$p_t$ and the regeneration which decreases with $p_t$ leads to a
relatively flat structure for central Cu+Cu collisions.

For the $p_t$\ dependence of $R_{AA}$ for central Au+Au
collisions, shown in the bottom panel of Fig.\ref{fig2}, while the
trend of the curves are similar to that for Cu+Cu collisions,
there are obvious features arisen from the stronger suppression
and stronger regeneration in Au+Au collisions. In comparison with
Cu+Cu collisions, the fireball formed in Au+Au collisions is much
hotter and larger and lasts much longer, and the initially
produced $J/\psi$s are strongly suppressed on their way out of the
fireball. Since not all the regenerated $J/\psi$s are from the
extremely hot medium, the effect of the stronger suppression in
Au+Au collisions on the regeneration is not so important as for
the initial production. Considering the fact that the regeneration
is proportional to the square of the number of binary collisions,
it is more important in central Au+Au collisions than in Cu+Cu
collisions. At low $p_t$ the regeneration contribution even
exceeds the initial production. It is the competition between the
stronger suppression of the initially produced $J/\psi$s and the
stronger regeneration at low $p_t$ that explains the overall
reduced $J/ \psi$ $R_{AA}$ and its minimum at $p_t\sim 2.5$ GeV/c
in Au+Au collisions. For central Cu+Cu collisions, there seems to
have a minimum too, see the top panel of Fig.\ref{fig2}. However,
due to the relatively small contribution from the regeneration,
our calculations do not show such a clear minimum.
\begin{figure}[!hbt]
\centering
\includegraphics[width=0.5\textwidth]{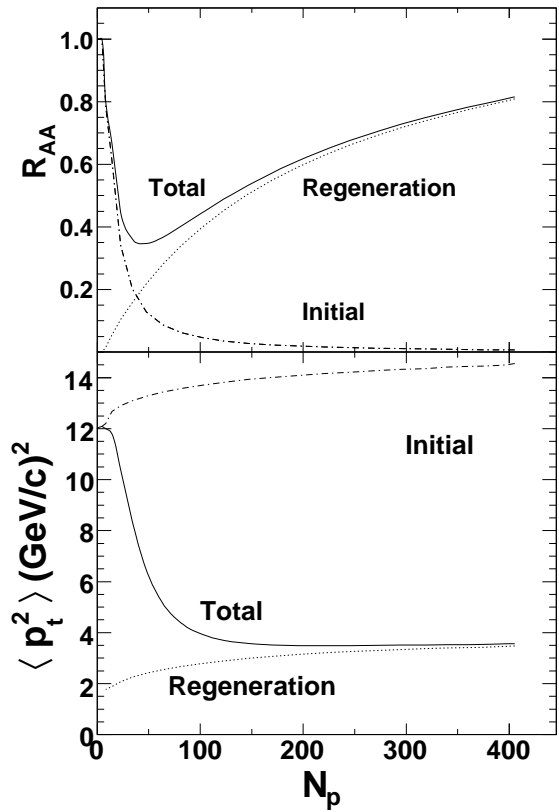}
\vspace{-0.5cm} \caption{The nuclear modification factor $R_{AA}$
and averaged transverse momentum square $\langle p_t^2\rangle$ as
function of $N_p$ in central rapidity region for Pb+Pb collisions
at LHC energy. Dot-dashed, dotted and solid lines represent
respectively the calculations with only initial production, only
regeneration and both contributions. } \label{fig3}
\end{figure}

As a prediction, we calculated the $J/\psi\ R_{AA}$ and $\langle
p_t^2\rangle$ at LHC energy, the results are shown in
Fig.\ref{fig3}. The input to the transport equation, namely the
charmonia and charm quark production cross sections in NN
collisions, is taken as $d\sigma_{NN}^{J/\psi}/dy=2\
\mu\textrm{b}$ and $d\sigma_{NN}^{c\bar c}/dy=0.7$ mb, estimated
by the CEM model and PYTHIA simulation~\cite{alice}. In comparison
with the heavy ion collisions at RHIC, the formed fireball at LHC
is hotter, bigger and longer lived, and almost all the initially
produced charmonia are eaten up by the medium. On the other hand,
there are much more charm quarks generated at LHC energy, the
regeneration controls the population in central and semi-central
collisions. This regeneration dominance leads to an increasing
$R_{AA}$ and a much stronger $p_t$ suppression at LHC for large
enough $N_p$. The rapid change of $R_{AA}$ and $\langle
p_t^2\rangle$ in the small $N_p$ region is due to the strong
competition between the initial production and regeneration, and
the degree of this competition decreases with increasing
centrality and ends when $N_p$ is large enough. The increasing
$R_{AA}$ and especially the small and saturated $\langle
p_t^2\rangle$ for central collisions can be regarded as signatures
of regeneration dominance at LHC, which are not expected in a
model with initial production only.

In summary, we investigated the transverse momentum dependence of
$J/\psi$ production in heavy ion collisions at RHIC and LHC
energies. By considering both initial production and regeneration
and solving the coupled hydrodynamic equations for medium
evolution and the transport equation for $J/\psi$ motion, we
focused on the centrality dependence of $\langle
p_t^2\rangle(N_p)$ and $p_t$ dependence of $R_{AA}(p_t)$ in Au+Au
and Cu+Cu collisions. While the high $p_t$ behavior is
characterized by the initial production, the regeneration and
initial production become equally important at low $p_t$. At RHIC,
the competition between the two production mechanisms leads to the
decrease of $J/\psi\ \langle
 p_t^2\rangle(N_p)$ in central collisions
and the minimum of $R_{AA}(p_t)$ at low $p_t$. It is necessary to
emphasize that the initial production alone cannot reproduce the
$p_t$ suppression and the minimum of $R_{AA}$ found at RHIC. At
LHC, almost all the initially produced charmonia are eaten up by
the hotter, larger and longer lived fireball, the $J/\psi$
behavior is dominated by the regeneration when $N_p$\ is large
enough. As a result, the $J/\psi\ R_{AA}$ increases with
centrality and decrease with $p_t$ in central collisions. On the
other hand, there is almost no regeneration at FAIR energy, and
the $J/\psi$ production is governed by the initial production. In
this case, both the $p_t$ suppression and the minimum of $R_{AA}$
will disappear in heavy ion collisions at FAIR energy.

{\underline{Acknowledgments:}} We are grateful to Xianglei Zhu and
Li Yan for their help in numerical calculations. The work is
supported by the NSFC grant No. 10735040, the 973-project No.
2007CB815000, and the U.S. Department of Energy under Contract No.
DE-AC03-76SF00098.

\end{document}